\documentclass[aps,twocolumn,showpacs,preprintnumbers,amsmath,amssymb,superscriptaddress,pre]{revtex4}
\usepackage[dvips]{graphicx}
\usepackage{dcolumn}
\usepackage{psfrag}
\usepackage{bm}
\usepackage{color}
\begin{document}

\preprint{APS/123-QED}

\title{Random walker in a temporally deforming higher-order potential forces observed in financial crisis}

\author{Kota Watanabe}
\email[]{watanabe@smp.dis.titech.ac.jp}
\affiliation{Department of Computational Intelligence \& Systems Science, Interdisciplinary Graduate 
      School of Science \& Engineering, Tokyo Institute of Technology, 4259-G3-52 Nagatsuta-cho,  
      Midori-ku, Yokohama 226-8502}

\author{Hideki Takayasu}
\affiliation{Sony Computer Science Laboratories,Inc., 3-14-13 Higashigotanda, Shinagawa-ku, Tokyo 
     141-0022}
\author{Misako Takayasu}
\affiliation{Department of Computational Intelligence \& Systems Science, Interdisciplinary Graduate 
      School of Science \& Engineering, Tokyo Institute of Technology, 4259-G3-52 Nagatsuta-cho,  
      Midori-ku, Yokohama 226-8502}     

\date{\today}
\begin{abstract}
Basic peculiarities of market price fluctuations are known to be well described by a recently developed random walk model in a temporally deforming quadric potential force whose center is given by a moving average of past price traces [Physica A 370, pp91-97, 2006]. By analyzing high-frequency financial time series of exceptional events such as bubbles and crashes, we confirm the appearance of nonlinear potential force in the markets. We show statistical significance of its existence by applying the information criterion. This new time series analysis is expected to be applied widely for detecting a non-stationary symptom in random phenomena.
\end{abstract}
\maketitle
\section{Introduction}
Financial bubbles and crashes have been occurring occasionally in nearly every market causing social troubles of various magnitudes. In the ordinary market state, prices fluctuate fairly randomly and directional prediction is almost impossible; however, during bubbles and crashes, the market prices are known to move quite asymmetrically \cite{mantegna}. Viewing the movement of prices in a coarse-grained way reveals that the growth of price in a bubble period is approximated either by an exponential function \cite{watanabe_PhysicaA382,watanabe_PhysicaA383}, a double exponential function \cite{mizuno_PhysicaA308}, or a function having a finite-time singularity \cite{sornette_PhysicaA294, sornette_JModPhysC13, ausloos_EPJB, ausloos_EPJB2, sornette_PhysicaA325, sornette_PhysicaA361}. In any case, it is a reasonable assumption that there are some special mechanisms of bubbles that are quite different from the ordinary market state. It is very important to develop the teqnique to quantify the financial risk of large price changes\cite{bouchaud}.

In the year 1900, Bachelier introduced the first random walk model as a model of market price fluctuations\cite{bachelier}. The stochastic model based on his mathematical theory was developed in the field of financial technology and it is widely accepted in the real financial market.

In the 1990s, analysis of high frequency tick-by-tick data by physicist clarified that market price is not a simple random walk\cite{stanley}. The market price has some empirically stylized facts which clearly deviate from a pure random walk. In order to build a model which fullfills those properties, many variants of random walk models of market prices have been proposed.

Recently, one of the authors (\textit{M.T.}) and her group introduced a new type of market model called potentials of unbalanced complex kinetics (PUCK), wherein the market price is described by a random walker in a temporally deforming potential force whose center is given by the trace of the walker \cite{takayasu_preprint, takayasu_PhysicaA370, alfi}. This model has quite different statistical properties from the case of the fixed potential function and its center, i.e., the Ulenbeck-Ornstein process\cite{uhlenbeck}. Moreover, the continuum limit of PUCK model is equivalent to the Langevin equation with time-dependent coefficients\cite{takayasu_PTP}.

From the viewpoint of this model with a quadric potential function, market states are categorized into five conditions, characterized by the potential function: (1) a pure random walk state, which is given by the case of no potential force. In mathematical finance, this condition is assumed to the generic property of the price fluctuation of a financial market; (2) a stable state, which is described by attractive potential force. Peinke and his group show existence of similar kind of attractive potential in foreign exchange market\cite{peinke}. Ausloos and his group applied the same method and obtained similar results for the stock index data\cite{ausloos_PRE}; (3) an unstable state which is described by a repulsive potential force; (4) a non-stationary state, which is characterized by a strong attractive force causing the exponential oscillation of price\cite{mizuno}; and (5) a non-stationary state, which is characterized by a strong repulsive force causing nearly monotonic price changes as a result. Data analysis based on PUCK has clarified that the market states are changing in various timescales.

The PUCK model is known to satisfy basic market statistics such as the power-law distribution of price changes\cite{mandelbrot, gabaix, gopi, farmer}, rapid decay of the autocorrelation of price changes, long correlation of the square of price changes\cite{stanley}, and abnormal diffusion properties\cite{plerou} in a short timescale \cite{takayasu_bmodel}. It has been clarified that the market potential force is closely related to the mass behavior of dealers, especially, the trend-following behavior \cite{yamada_EPJB}.  Morever, the Nobel prize laureled market model - the autoregressive conditional heteroskedasticity (ARCH) model - is derived in a very special limit case, where a nonlinear potential force appears with randomly chosen signs at each time step \cite{takayasu_PhysicaA383}. 

In this paper, we reformulate the data analysis method of PUCK taking into account the higher-order potential functions, and we focus our attention on finding precursors of bubbles and crashes. The data we analyze here as an example is the tick-by-tick data of yen-dollar exchange rates, which showed the largest rate change in 1998. 
\section{The PUCK model}
\begin{figure}
\psfrag{dp1(tn)}[8][]{\scriptsize $\Delta P_{1}(t_{n})$}
\psfrag{tn}[8][]{\scriptsize $t_{n}$}
\psfrag{dp(tn+dt)}[8][]{\scriptsize $\Delta P_{2}(t_{n}+1)$}
\psfrag{tn+dt}[8][]{\scriptsize $t_{n}+1$}
\psfrag{P(t+1)-P(t)}[8][]{\scriptsize $P(t+1)-P(t)$}
\psfrag{P(t)-PM(t)}[8][]{\scriptsize $P(t)-P_{M}(t)$}
\psfrag{P(t)}[8][]{\scriptsize $P(t)$}
\psfrag{PM(t)}[8][]{\scriptsize $P_{M}(t)$}
\includegraphics[scale=.36]{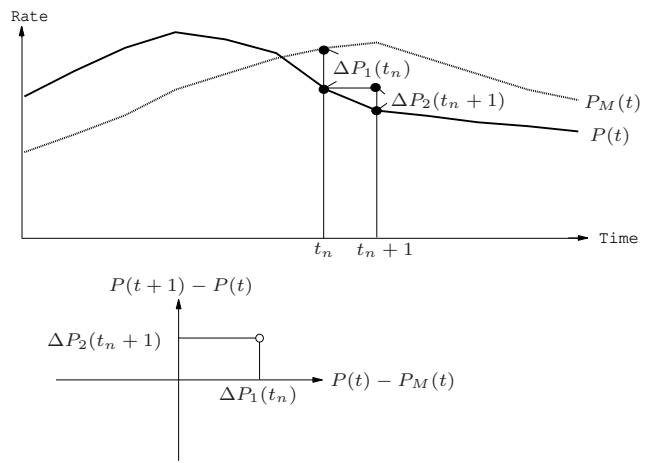}
\caption{Schematic diagram of the PUCK model. For given market time series we calculate $P(t+1)-P(t)$ and $P(t)-P_{M}(t)$ as shown in the upper diagram. The plots in the lower diagram is proportional to the derivative of the potential function, $-\frac{d}{dp}U(p, t)$.}
\label{concept}
\end{figure}

The PUCK model is represented by the following set of equations. \begin{eqnarray}
P(t+1)-P(t)&=&-\frac{d}{dp}U(p, t)|_{p=P(t)-P_{M}(t)} \nonumber \\
            &&+f(t) \label{linearmodel} \\
P_{M}(t)&=&\frac{1}{M} \displaystyle \sum_{k=0}^{M-1}P(t-k)
\end{eqnarray}
Here, $P(t)$ is the noise-reduced market price at the t-th tick obtained by applying the optimal moving average \cite{ohnishi_PhysicaA344}, $f(t)$ is the random noise typically a Gaussian white noise, $U(p, t)$ is the potential function which is approximated by the following Taylor expansion form with time-dependent coefficients $b_{k}(t)$,
\begin{equation}
U(p, t)=\displaystyle \sum_{k=1}^{\infty}\frac{b_{k}(t)}{k}p^{k}. \label{potentialfunction}
\end{equation}
$P_{M}(t)$ is the average of the past price changes with size M ticks, and it is assumed that this point gives the center of the potential function that moves with the random walker's footprint (Fig. \ref{concept}). For the yen-dollar market a typical value of M is smaller than ten. We postulate that it is an averaged expectation price of the market dealers.
\par
\begin{figure*}
\psfrag{U}[]{\scriptsize $U$}
\psfrag{P(t)-PM(t)}[1][]{\scriptsize    $P(t)-P_{M}(t)$}
\includegraphics[scale=.35]{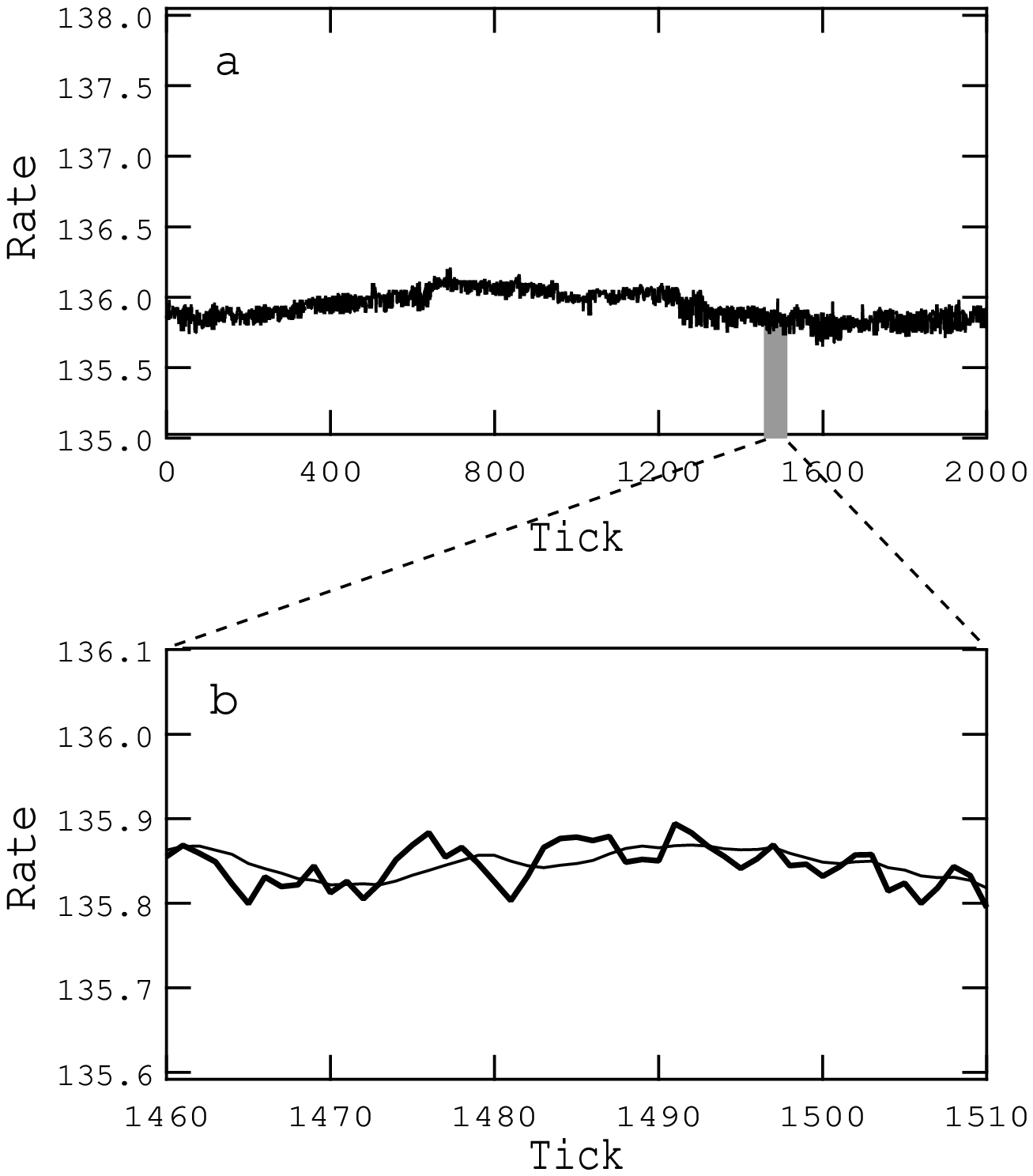}
\includegraphics[scale=.32]{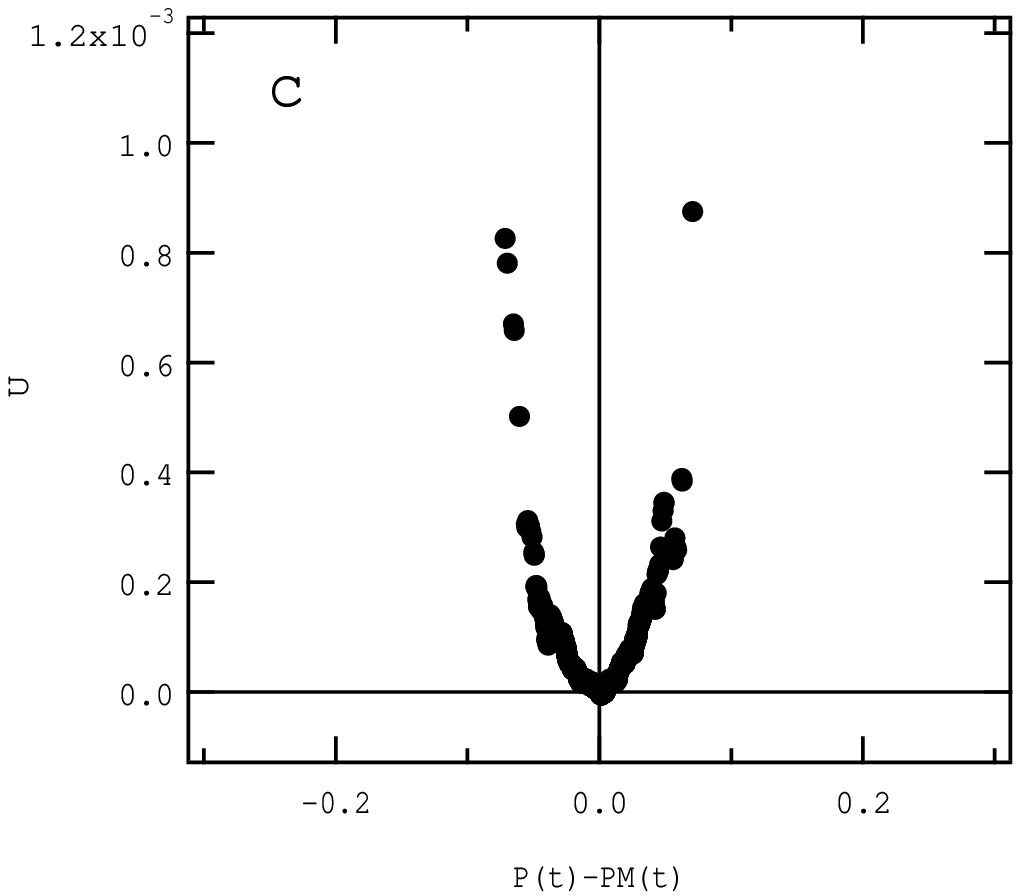}
\includegraphics[scale=.35]{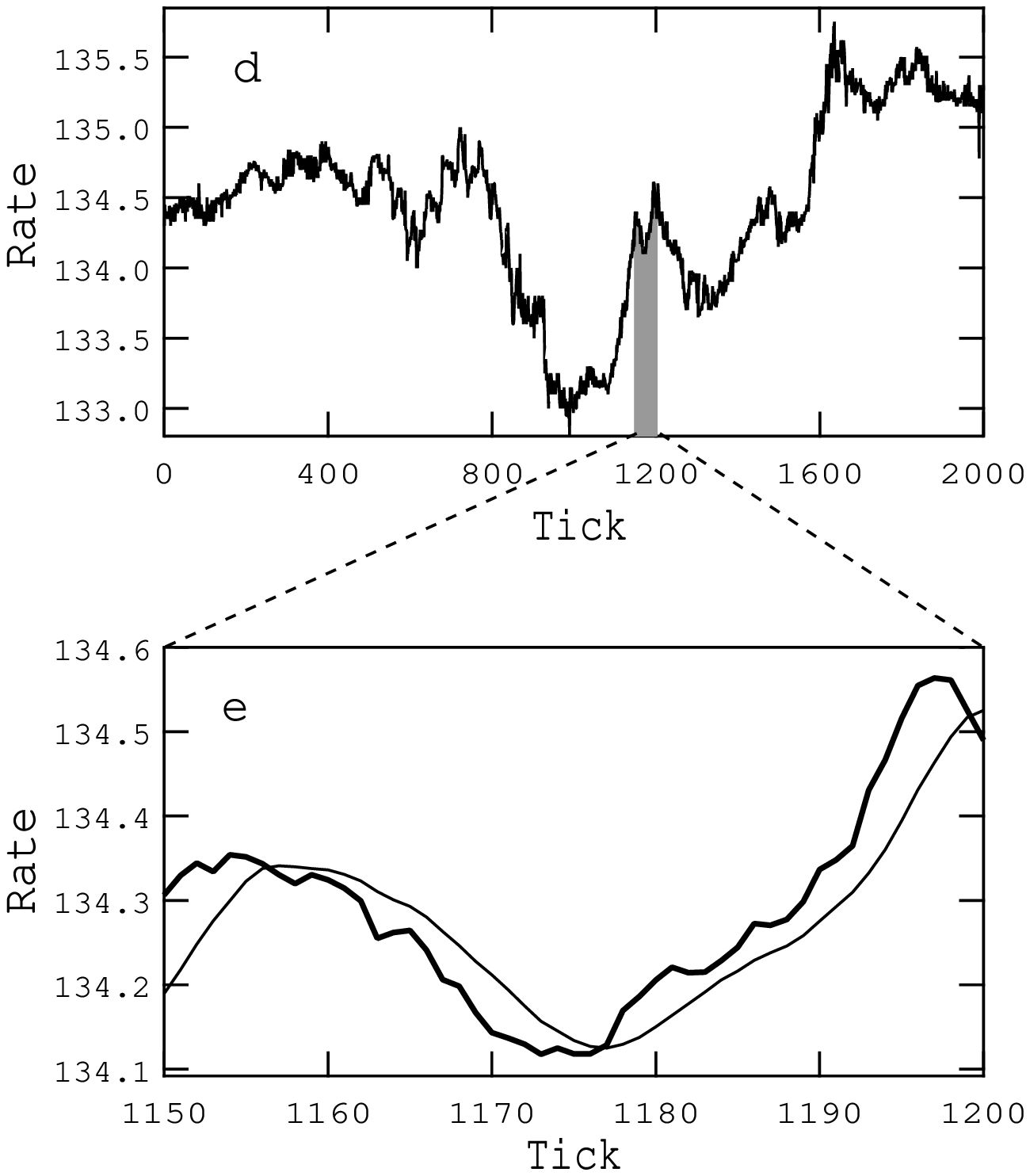}
\includegraphics[scale=.32]{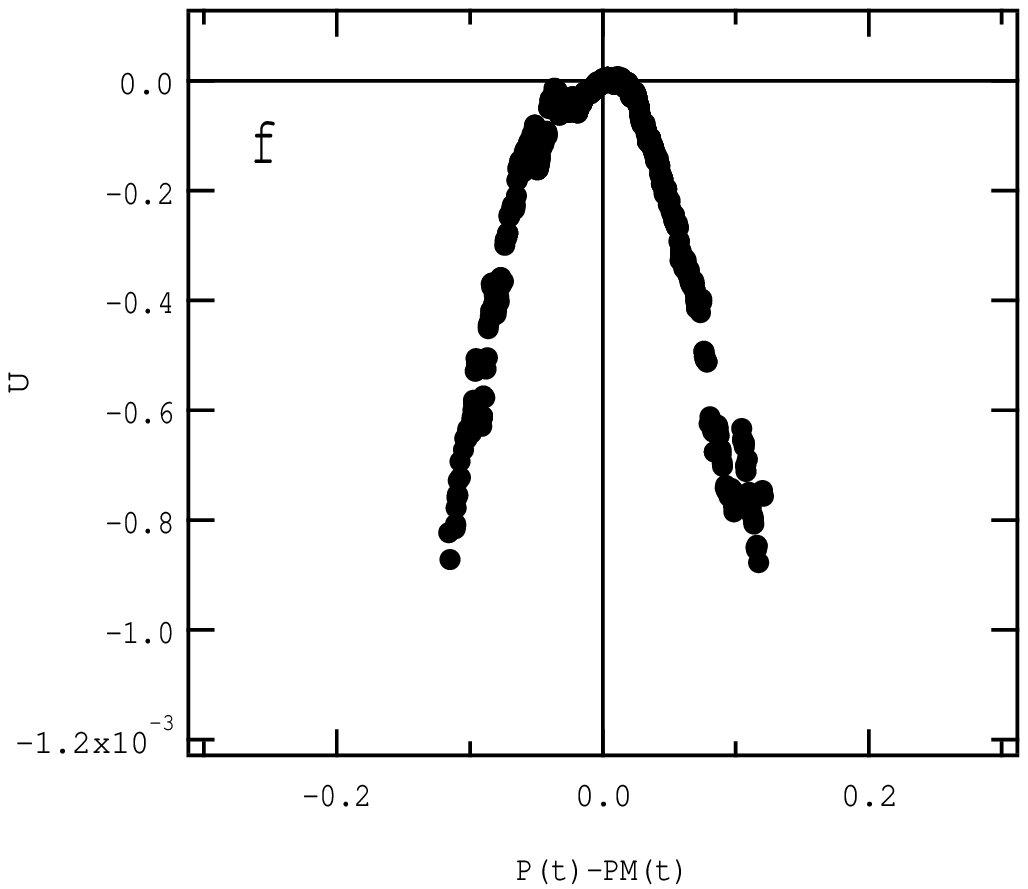}
\caption{Examples of estimated potential functions. For given yen-dollar rates, Fig. \ref{example}a and Fig. \ref{example}d, the relations between $P(t)$ (heavy line) and $P_{M}(t)$ (thin line) are observed as shown in Fig. \ref{example}b and Fig. \ref{example}e, respectively. The potential functions, Fig. \ref{example}c and Fig. \ref{example}f, are estimated by integrating the plots of the lower diagram of Fig. \ref{concept}}.
\label{example}
\end{figure*}

In Fig. \ref{example} we show typical examples of potential functions observed in the yen-dollar market for given 2000 data points. Here, the mean interval of tick is about fifteen seconds in both cases. The market fluctuations of Fig. \ref{example}a and Fig. \ref{example}d look very different intuitively; however, the square of the market price fluctuations in unit time, called the volatility, takes about the same value in both cases. By applying the PUCK model we can clearly observe the difference in the quadric term of the potential function as shown in Fig. \ref{example}c and Fig. \ref{example}f. In the case of Fig. \ref{example}c the value of $b_{2}(t)$ is positive and the market fluctuation is stable, while in the case of Fig. \ref{example}f the value of $b_{2}(t)$ is negative and market fluctuation is unstable. As typically shown here the potential coefficients other than $b_{2}(t)$, such as $b_{1}(t)$ or $b_{3}(t)$, are generally negligibly small in ordinary market states.
It is found that these quadric potential functions can generally be found in ordinary states of financial markets; however, in particular periods, when prices move rather monotonically we can observe a new type of nonquadric potential functions such as a cubic or a quartic function, as shown later. 

In order to describe such nonlinear potential functions correctly, we introduce a higher-order term in the potential model as follows.         
\begin{equation}
U(p,t)=\frac{b_{1}(t)}{2}p^{2}+\frac{b_{\gamma}}{\gamma+1}p^{\gamma+1} \label{nonlinearmodel}
\end{equation}
This higher-order potential model can reproduce all kinds of typical price motions in bubbles, crashes and hyper-inflations as mentioned in the introduction of this paper\cite{takayasu_others}.

\begin{figure}
\psfrag{U}[8][]{$U$}
\psfrag{P(t)-PM(t)}[8][]{\scriptsize $P(t)-P_{M}(t)$}
\includegraphics[scale=.35]{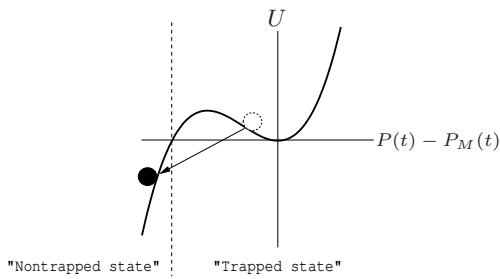}
\caption{Schematic diagram of the transition of price changes from a stable trapped state to a highly unstable nontrapped state.}
\label{qubic_concept}
\end{figure}

When the market potential function is described by a cubic function with a local minimum point, there are obviously two states for the random walker as shown schematically in Fig. \ref{qubic_concept}. The first state is the trapped state that is practically the same as the case of a random walk in a stable quadric potential function, however, with a finite possibility the walker goes over the potential barrier. The second state is the nontrapped state, wherein the walker goes down the potential slope indefinitely following a double-exponential growth or even the finite time divergence in the continuum limit. In particular, we can expect a sudden transition of the movement of prices from a stable state to a highly unstable state even when the market potential function is invariant. 
\section{Estimation of non-linear potential functions}
For estimating the nonlinear potential function from real data of bubbles or crashes, we need to consider the meaning of probability distribution of the independent random variable $f(t)$ in Eq. (\ref{linearmodel}). From a skeptical viewpoint about the existence of the potential forces, one might think that any price movement can be fully described by only the random noise term $f(t)$, assuming that $f(t)$ takes extremely asymmetric values repeatedly by chance. This case is theoretically realizable; however, the probability of the occurrence of the whole event, which is estimated by the products of probability density of $f(t)$, becomes negligibly small for the cases of bubbles or crashes. This is confirmed systematically in the following manner. 

For the given symmetric probability distribution of $f(t)$, $w(f(t))$, we can calculate the probability of the occurrence of any time series by assuming that the process is governed by Eq. (\ref{linearmodel}) with the potential function given by Eq. (\ref{nonlinearmodel}), and the parameters of the model are constants, namely $b_{1}(t)=b_{1},b_{\gamma}(t)=b_{\gamma}$, in the period from $t = n$ to $n + N$. The probability of the occurrence or the likelihood of a given market price time series, $\{P(n), P(n+1),\cdots, P(n+N)\}$, is calculated by the following equation.
\begin{equation}
l(b_{1}, \gamma, b_{\gamma}, M)=\prod_{t=n}^{n+N}w(f(t)) \label{likelihood} 
\end{equation}
where each value of $f(t)$ is determined from Eq. (\ref{linearmodel}) with Eq. (\ref{nonlinearmodel}). As for the functional form of the probability density of $f(t)$, we apply the Gaussian distribution with zero mean. 

In statistics, this type of maximum-likelihood method is very popular, and the method of choosing the optimum number of parameters has already been established. In general, a model with more parameters has a greater tendency to receive a higher value of likelihood. In order to solve an overfitting problem and to evaluate which of the two models with different number of parameters is better, we should not use the likelihood itself, but apply the information criterion called Akaike information criterion (AIC) that is composed of the logarithm of likelihood and the term of penalty, which depends on the number of parameters, $k$ \cite{akaike_1973,schwarz_BIC}.
\begin{equation}
\text{AIC}=-2\ln l(b_{1}, \gamma, b_{\gamma}, M)+2k \label{AIC} 
\end{equation}

\begin{figure}
\psfrag{b1}[8][]{\scriptsize $b_{1}$}
\psfrag{b2}[8][]{\scriptsize $b_{2}$}
\psfrag{M}[8][]{\scriptsize $M$}
\includegraphics[scale=.34]{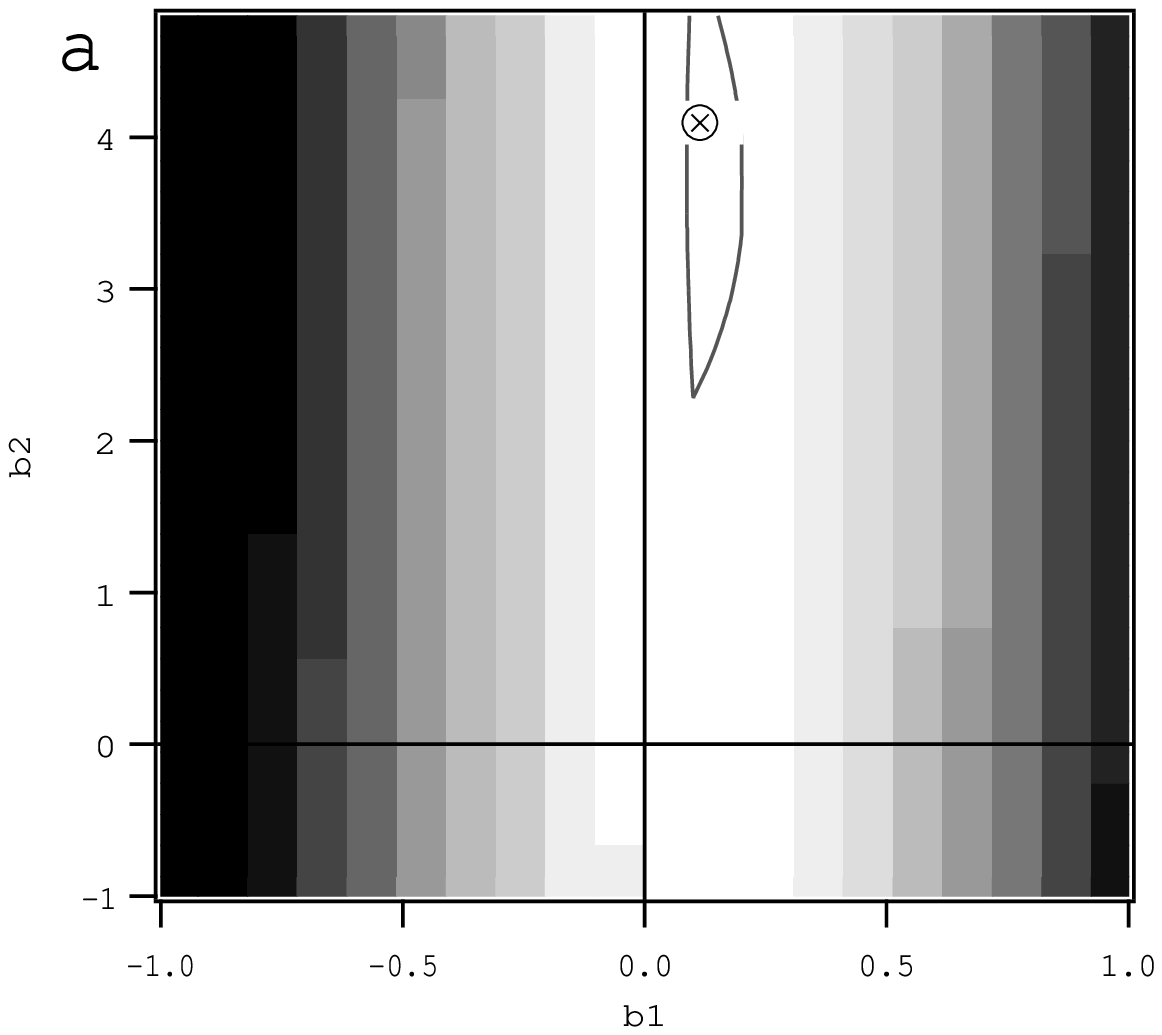}
\includegraphics[scale=.34]{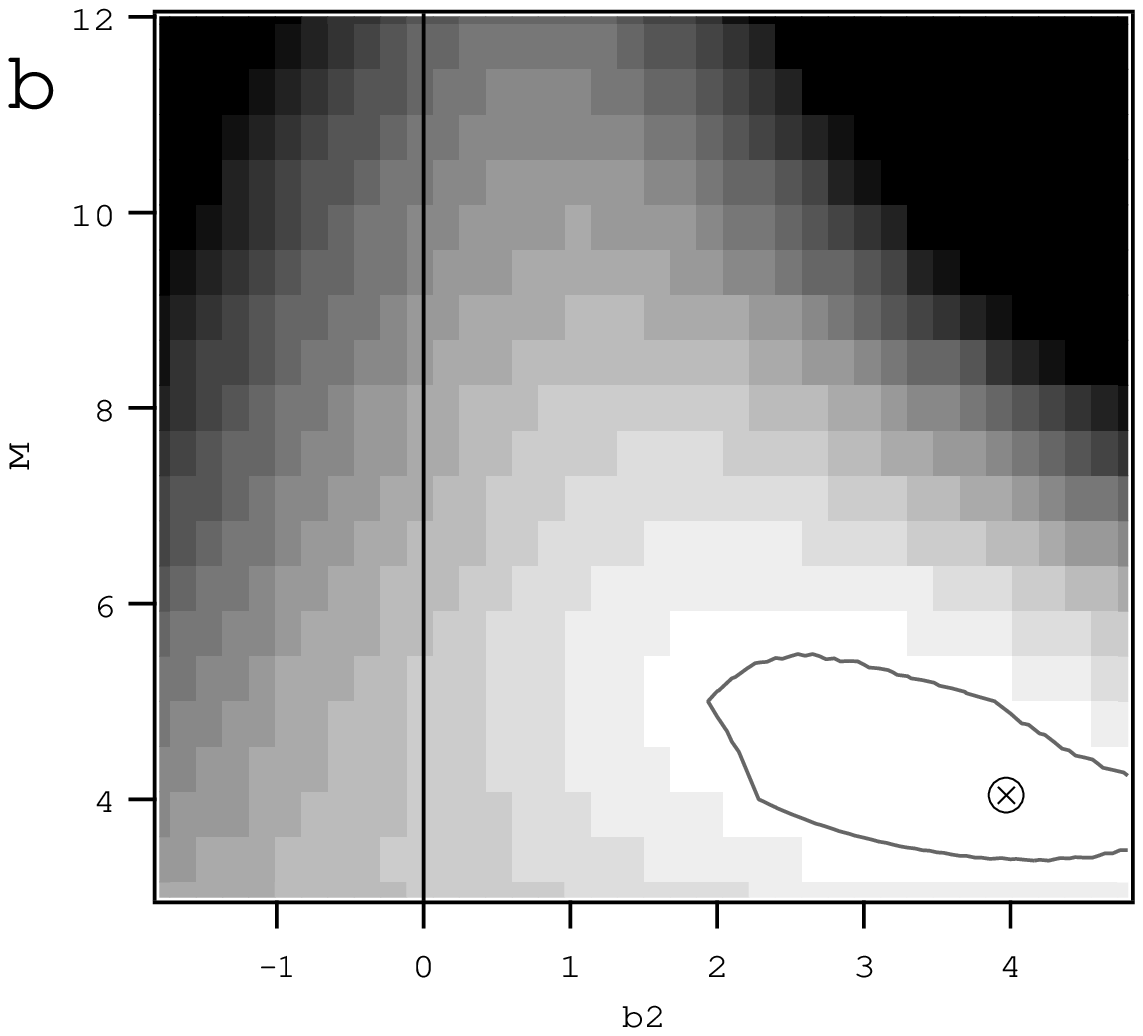}
\caption{Parameter space of the values of AIC and the optimal parameters ($\otimes$) in the case of Fig. \ref{yd1998_1}-I\hspace{-.1em}I. $b_{1}$ vs $b_{2}$ (a), $M$ vs $b_{2}$ (b)}
\label{MLE_yd98}
\end{figure}

\begin{figure*}
\includegraphics[scale=.37]{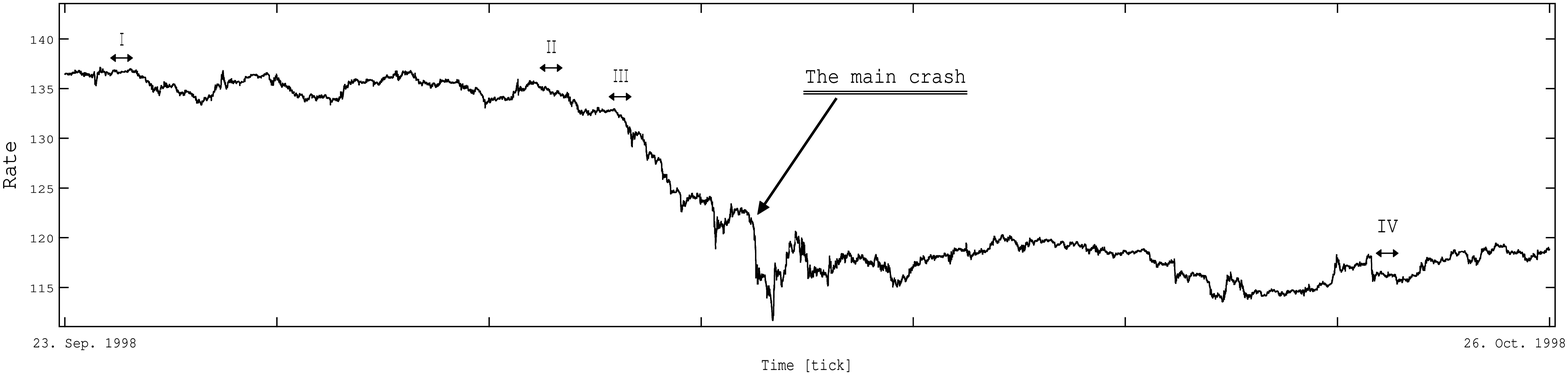}\\
\psfrag{U}[8][]{\scriptsize $U$}
\psfrag{P(t)-PM(t)}[8][]{\scriptsize $P(t)-P_{M}(t)$}
\includegraphics[scale=.38]{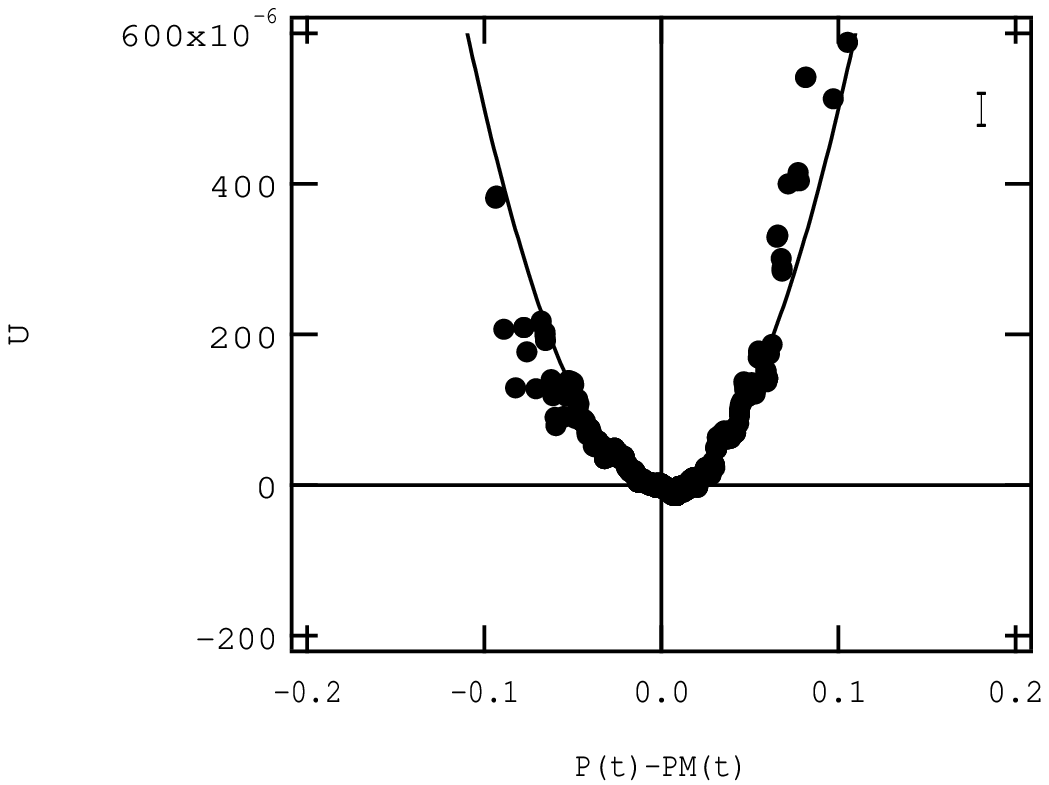}
\includegraphics[scale=.38]{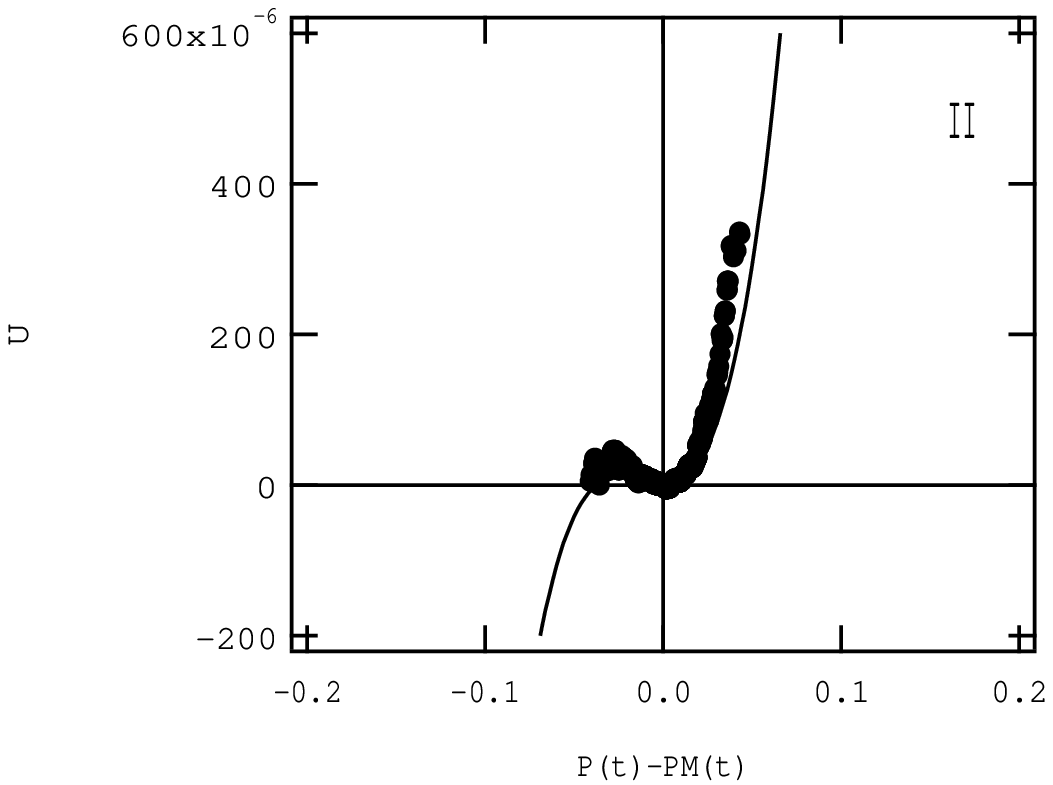}
\includegraphics[scale=.38]{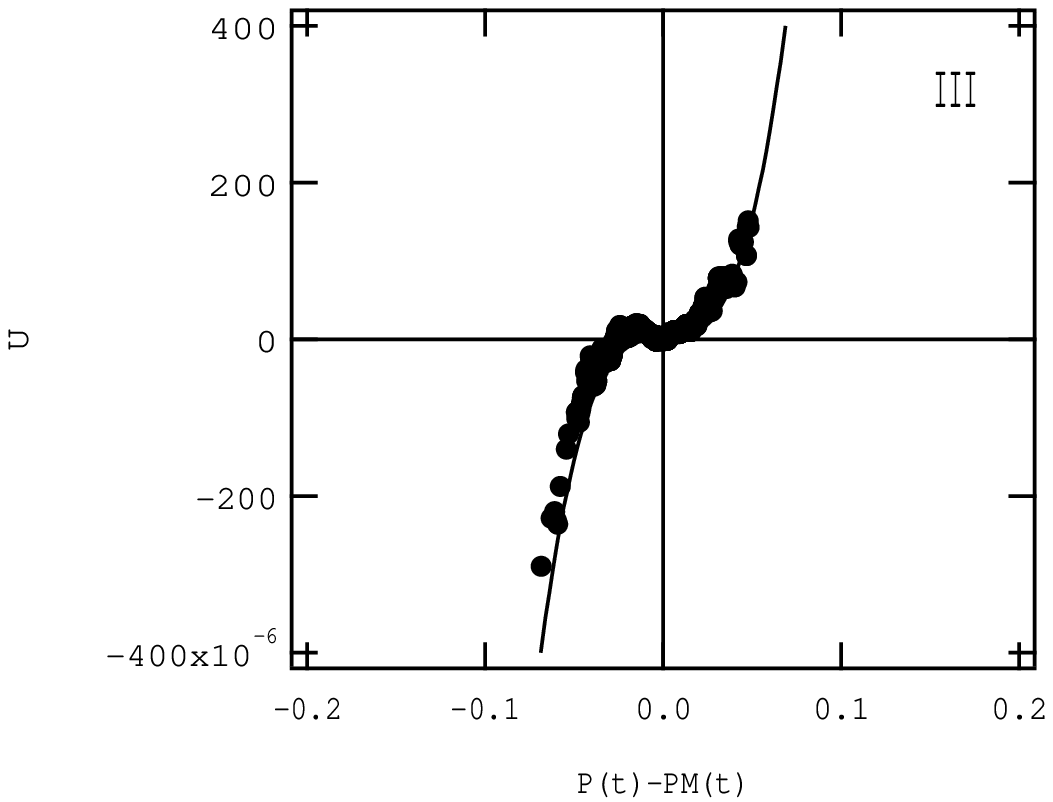}
\includegraphics[scale=.38]{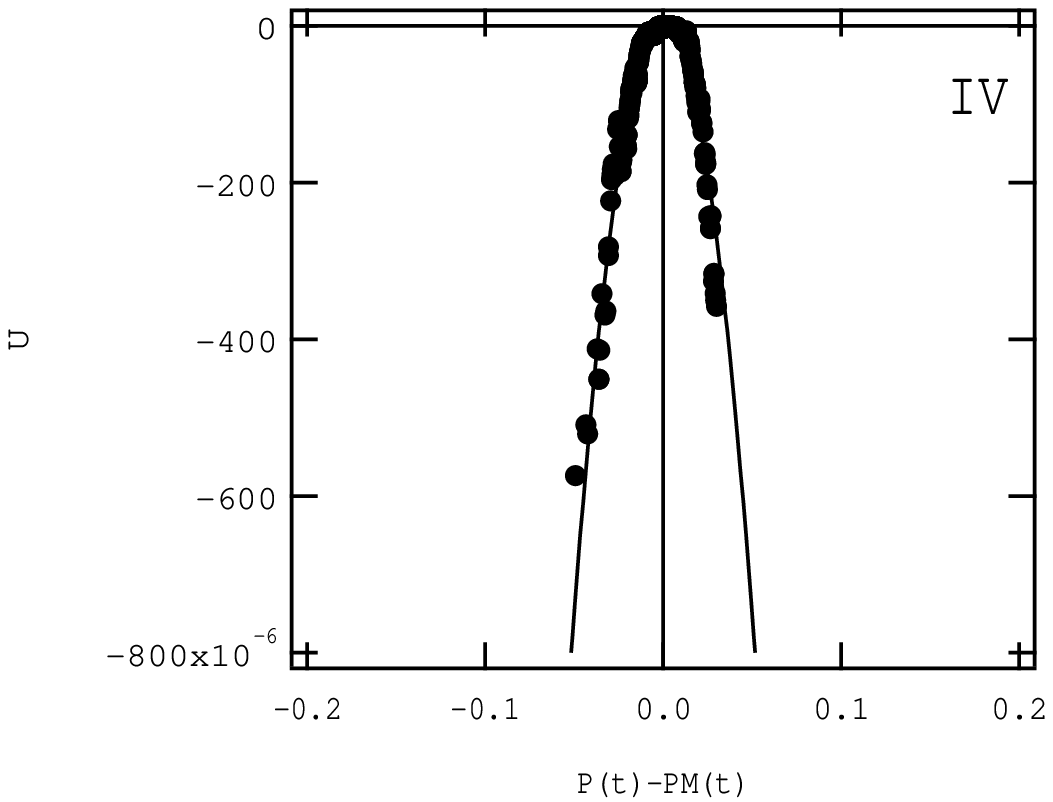}
\caption{Estimated optimal potential forms by the minimum information criterion procedure in the yen-dollar exchange rates of 1998. Potential functions are estimated for periods shown in I, I\hspace{-.1em}I, I\hspace{-.1em}I\hspace{-.1em}I, and I\hspace{-.1em}V. In these figures, we show directly observed potential functions ($\bullet$), the optimal potential forms assuming the normal distribution (thin line) to the term of $f(t)$.}
\label{yd1998_1}
\end{figure*}

Using this method, the contour plot of the value of AIC for a given time series (period II in Fig. \ref{yd1998_1}) is plotted in Fig. \ref{MLE_yd98}. Here, the control parameters for determining the center of potential force are $b_{1}$; $\gamma$; $b_{\gamma}$ and $M$, and the value of the moving average size $M$. The case of no potential force corresponds to the parameter values $b_{1}=0$, and $b_{\gamma}=0$. From these figures we can find a non-trivial set of parameters that maximize the likelihood. 

We can introduce, for example, one more nonlinear term in Eq. (\ref{nonlinearmodel}) and compare the value of AIC with that of one nonlinear term model. As a result, the best market price model in view of the AIC value is given by the model already introduced in Eq. (\ref{nonlinearmodel}) with the lowest integer nonlinear exponent $\gamma = 2$, that is, the cubic potential function. We also apply the Bayesian information criterion (BIC) and replace the distribution of $f(t)$ by a distribution with long tails, but the results are almost the same in any case. 

In Fig. \ref{yd1998_1}, we illustrate a part of the yen-dollar exchange rates from September 23rd to October 26th of 1998, during this period the yen-doller market showed historically the largest fluctuations. In period I, the potential function is quadratic and stable. In period II, the left side of the potential function becomes shorter and the best fit cubic function indicates the possibility that the exchange rate might crash. In period III, we can clearly observe the cubic potential with a very shallow well, which indicates that the exchange rate may tend to decrease monotonically. In period IV, the main crash is over, however, the potential function is unstable and symmetric, which implies that the market will be turbulent, but the fluctuations will be symmetric. 

We confirmed that these results do not change even if we replace the distribution function of $f(t)$ from the normal distribution to a fat-tailed distribution or to an asymmetric distribution. It should be noted that we find that the potential function becomes cubic a little before the largest crash, which occurs in the middle of the figure. This indicates that the cubic potential function can be a precursor to the largest event.
\section{Conclusion}
In this paper, we introduced a higher-order potential model suitable for the analysis of the market forces observed in financial bubbles or crashes. In order to show the statistical significance of the existence of higher-order potential functions, we applied the information criterion. In the given example, a clear cubic potential function was detected before the major crash, which can be considered as a precursor to a devastating event. Similar nonlinear potential functions can be confirmed in many other examples of bubbles and crashes of financial markets.

It should be noted that we need to refine the algorithm for quicker estimation of the higher-order potential function. This is needed because there is a finite possibility that the cubic potential disappears shortly without transition to the unstable state, in that case no drastic event occurs as a result. Another reason is that a crash or a bubble actually occurs; however, there is a possibility that the form of potential function changes quickly resulting a sharp market rebound.

One thing is for certain that the existence of a cubic potential function is a precursor of a big risk that might be encountered in the near future. By properly introducing risk aversion actions on the basis of these results, we might be able to avoid catastrophic financial events in the near future, though the course of actions is not yet well established. It is an urgent task to clarify the origin of such higher-order market potential force from the viewpoint of human activity.

Application of the method we developed in this paper is not limited to financial market data. As known from the formulation this method is applicable to any time series showing random-walk-like behaviors. For examples, medical data from patients, machine condition data in a factory, environmental and climate data are promising candidates for applying this method for early detection of non-stationary symptoms. 
\section{Acknowledgement}
This work is partly supported by Research Fellowships of the Japan Society for the Promotion of Science for Young Scientists (K.W.) and Japan Society for the Promotion of Science, Grant-in-Aid for Scientific Research $ \sharp $ 16540346 (M.T.).


\begin{thebibliography}{33}
\expandafter\ifx\csname natexlab\endcsname\relax\def\natexlab#1{#1}\fi
\expandafter\ifx\csname bibnamefont\endcsname\relax
  \def\bibnamefont#1{#1}\fi
\expandafter\ifx\csname bibfnamefont\endcsname\relax
  \def\bibfnamefont#1{#1}\fi
\expandafter\ifx\csname citenamefont\endcsname\relax
  \def\citenamefont#1{#1}\fi
\expandafter\ifx\csname url\endcsname\relax
  \def\url#1{\texttt{#1}}\fi
\expandafter\ifx\csname urlprefix\endcsname\relax\def\urlprefix{URL }\fi
\providecommand{\bibinfo}[2]{#2}
\providecommand{\eprint}[2][]{\url{#2}}

\bibitem[{\citenamefont{Lillo and Mantegna}(2000)}]{mantegna}
\bibinfo{author}{\bibfnamefont{F.}~\bibnamefont{Lillo}} \bibnamefont{and}
  \bibinfo{author}{\bibfnamefont{R.~N.} \bibnamefont{Mantegna}},
  \bibinfo{journal}{The European Physical Journal B}
  \textbf{\bibinfo{volume}{15}}, \bibinfo{pages}{603} (\bibinfo{year}{2000}).

\bibitem[{\citenamefont{Watanabe
  et~al.}(2007{\natexlab{a}})\citenamefont{Watanabe, Takayasu, and
  Takayasu}}]{watanabe_PhysicaA382}
\bibinfo{author}{\bibfnamefont{K.}~\bibnamefont{Watanabe}},
  \bibinfo{author}{\bibfnamefont{H.}~\bibnamefont{Takayasu}}, \bibnamefont{and}
  \bibinfo{author}{\bibfnamefont{M.}~\bibnamefont{Takayasu}},
  \bibinfo{journal}{Physica A} \textbf{\bibinfo{volume}{382}},
  \bibinfo{pages}{336} (\bibinfo{year}{2007}{\natexlab{a}}).

\bibitem[{\citenamefont{Watanabe
  et~al.}(2007{\natexlab{b}})\citenamefont{Watanabe, Takayasu, and
  Takayasu}}]{watanabe_PhysicaA383}
\bibinfo{author}{\bibfnamefont{K.}~\bibnamefont{Watanabe}},
  \bibinfo{author}{\bibfnamefont{H.}~\bibnamefont{Takayasu}}, \bibnamefont{and}
  \bibinfo{author}{\bibfnamefont{M.}~\bibnamefont{Takayasu}},
  \bibinfo{journal}{Physica A} \textbf{\bibinfo{volume}{383}},
  \bibinfo{pages}{120} (\bibinfo{year}{2007}{\natexlab{b}}).

\bibitem[{\citenamefont{Mizuno et~al.}(2002)\citenamefont{Mizuno, Takayasu, and
  Takayasu}}]{mizuno_PhysicaA308}
\bibinfo{author}{\bibfnamefont{T.}~\bibnamefont{Mizuno}},
  \bibinfo{author}{\bibfnamefont{H.}~\bibnamefont{Takayasu}}, \bibnamefont{and}
  \bibinfo{author}{\bibfnamefont{M.}~\bibnamefont{Takayasu}},
  \bibinfo{journal}{Physica A} \textbf{\bibinfo{volume}{308}},
  \bibinfo{pages}{411} (\bibinfo{year}{2002}).

\bibitem[{\citenamefont{Johansen and Sornette}(2001)}]{sornette_PhysicaA294}
\bibinfo{author}{\bibfnamefont{A.}~\bibnamefont{Johansen}} \bibnamefont{and}
  \bibinfo{author}{\bibfnamefont{D.}~\bibnamefont{Sornette}},
  \bibinfo{journal}{Physica A} \textbf{\bibinfo{volume}{294}},
  \bibinfo{pages}{465} (\bibinfo{year}{2001}).

\bibitem[{\citenamefont{Sornette and Andersen}(2002)}]{sornette_JModPhysC13}
\bibinfo{author}{\bibfnamefont{D.}~\bibnamefont{Sornette}} \bibnamefont{and}
  \bibinfo{author}{\bibfnamefont{J.}~\bibnamefont{Andersen}},
  \bibinfo{journal}{Int. J. Mod. Phys. C} \textbf{\bibinfo{volume}{13}},
  \bibinfo{pages}{171} (\bibinfo{year}{2002}).

\bibitem[{\citenamefont{Vandewalle et~al.}(1998)\citenamefont{Vandewalle,
  Ausloos, Boveroux, and Minguet}}]{ausloos_EPJB}
\bibinfo{author}{\bibfnamefont{N.}~\bibnamefont{Vandewalle}},
  \bibinfo{author}{\bibfnamefont{M.}~\bibnamefont{Ausloos}},
  \bibinfo{author}{\bibfnamefont{P.}~\bibnamefont{Boveroux}}, \bibnamefont{and}
  \bibinfo{author}{\bibfnamefont{A.}~\bibnamefont{Minguet}},
  \bibinfo{journal}{The European Physical Journal B}
  \textbf{\bibinfo{volume}{4}}, \bibinfo{pages}{139} (\bibinfo{year}{1998}).

\bibitem[{\citenamefont{Vandewalle et~al.}(1999)\citenamefont{Vandewalle,
  Ausloos, Boveroux, and Minguet}}]{ausloos_EPJB2}
\bibinfo{author}{\bibfnamefont{N.}~\bibnamefont{Vandewalle}},
  \bibinfo{author}{\bibfnamefont{M.}~\bibnamefont{Ausloos}},
  \bibinfo{author}{\bibfnamefont{P.}~\bibnamefont{Boveroux}}, \bibnamefont{and}
  \bibinfo{author}{\bibfnamefont{A.}~\bibnamefont{Minguet}},
  \bibinfo{journal}{The European Physical Journal B}
  \textbf{\bibinfo{volume}{9}}, \bibinfo{pages}{355} (\bibinfo{year}{1999}).

\bibitem[{\citenamefont{Sornette et~al.}(2003)\citenamefont{Sornette, Takayasu,
  and Zhou}}]{sornette_PhysicaA325}
\bibinfo{author}{\bibfnamefont{D.}~\bibnamefont{Sornette}},
  \bibinfo{author}{\bibfnamefont{H.}~\bibnamefont{Takayasu}}, \bibnamefont{and}
  \bibinfo{author}{\bibfnamefont{W.~X.} \bibnamefont{Zhou}},
  \bibinfo{journal}{Physica A} \textbf{\bibinfo{volume}{325}},
  \bibinfo{pages}{492} (\bibinfo{year}{2003}).

\bibitem[{\citenamefont{Zhou and Sornette}(2006)}]{sornette_PhysicaA361}
\bibinfo{author}{\bibfnamefont{W.~X.} \bibnamefont{Zhou}} \bibnamefont{and}
  \bibinfo{author}{\bibfnamefont{D.}~\bibnamefont{Sornette}},
  \bibinfo{journal}{Physica A} \textbf{\bibinfo{volume}{361}},
  \bibinfo{pages}{297} (\bibinfo{year}{2006}).

\bibitem[{\citenamefont{Bouchaud and Potters}(2000)}]{bouchaud}
\bibinfo{author}{\bibfnamefont{J.~P.} \bibnamefont{Bouchaud}} \bibnamefont{and}
  \bibinfo{author}{\bibfnamefont{M.}~\bibnamefont{Potters}},
  \emph{\bibinfo{title}{Theory of financial risks}}
  (\bibinfo{publisher}{Cambridge University Press},
  \bibinfo{address}{Cambridge}, \bibinfo{year}{2000}).

\bibitem[{\citenamefont{Bachelier}(1964)}]{bachelier}
\bibinfo{author}{\bibfnamefont{L.}~\bibnamefont{Bachelier}},
  \emph{\bibinfo{title}{The Random Character of Stock Market Prices}}
  (\bibinfo{publisher}{The MIT Press}, \bibinfo{address}{Cambridge},
  \bibinfo{year}{1964}).

\bibitem[{\citenamefont{Mantegna and Stanley}(2000)}]{stanley}
\bibinfo{author}{\bibfnamefont{R.}~\bibnamefont{Mantegna}} \bibnamefont{and}
  \bibinfo{author}{\bibfnamefont{H.~E.} \bibnamefont{Stanley}},
  \emph{\bibinfo{title}{An Introduction to Econophysics: Correlation and
  Complexity in Finance}} (\bibinfo{publisher}{Cambridge University Press},
  \bibinfo{address}{Cambridge}, \bibinfo{year}{2000}).

\bibitem[{\citenamefont{Takayasu et~al.}({\natexlab{a}})\citenamefont{Takayasu,
  Mizuno, and Takayasu}}]{takayasu_preprint}
\bibinfo{author}{\bibfnamefont{M.}~\bibnamefont{Takayasu}},
  \bibinfo{author}{\bibfnamefont{T.}~\bibnamefont{Mizuno}}, \bibnamefont{and}
  \bibinfo{author}{\bibfnamefont{H.}~\bibnamefont{Takayasu}},
  \eprint{physics/0509020}.

\bibitem[{\citenamefont{Takayasu et~al.}(2006)\citenamefont{Takayasu, Mizuno,
  and Takayasu}}]{takayasu_PhysicaA370}
\bibinfo{author}{\bibfnamefont{M.}~\bibnamefont{Takayasu}},
  \bibinfo{author}{\bibfnamefont{T.}~\bibnamefont{Mizuno}}, \bibnamefont{and}
  \bibinfo{author}{\bibfnamefont{H.}~\bibnamefont{Takayasu}},
  \bibinfo{journal}{Physica A} \textbf{\bibinfo{volume}{370}},
  \bibinfo{pages}{91} (\bibinfo{year}{2006}).

\bibitem[{\citenamefont{Alfi et~al.}(2006)\citenamefont{Alfi, Coccetti,
  Marotta, Pietronero, and Takayasu}}]{alfi}
\bibinfo{author}{\bibfnamefont{V.}~\bibnamefont{Alfi}},
  \bibinfo{author}{\bibfnamefont{F.}~\bibnamefont{Coccetti}},
  \bibinfo{author}{\bibfnamefont{M.}~\bibnamefont{Marotta}},
  \bibinfo{author}{\bibfnamefont{L.}~\bibnamefont{Pietronero}},
  \bibnamefont{and} \bibinfo{author}{\bibfnamefont{M.}~\bibnamefont{Takayasu}},
  \bibinfo{journal}{Physica A} \textbf{\bibinfo{volume}{370}},
  \bibinfo{pages}{30} (\bibinfo{year}{2006}).

\bibitem[{\citenamefont{Uhlenbeck and Ornstein}(1930)}]{uhlenbeck}
\bibinfo{author}{\bibfnamefont{G.}~\bibnamefont{Uhlenbeck}} \bibnamefont{and}
  \bibinfo{author}{\bibfnamefont{L.}~\bibnamefont{Ornstein}},
  \bibinfo{journal}{Physical Review.} \textbf{\bibinfo{volume}{36}},
  \bibinfo{pages}{823} (\bibinfo{year}{1930}).

\bibitem[{\citenamefont{Takayasu and Takayasu}()}]{takayasu_PTP}
\bibinfo{author}{\bibfnamefont{M.}~\bibnamefont{Takayasu}} \bibnamefont{and}
  \bibinfo{author}{\bibfnamefont{H.}~\bibnamefont{Takayasu}},
  \bibinfo{note}{submitted to Progress of Theoretical Physics}.

\bibitem[{\citenamefont{Friedrich et~al.}(2000)\citenamefont{Friedrich, Peinke,
  and Renner}}]{peinke}
\bibinfo{author}{\bibfnamefont{R.}~\bibnamefont{Friedrich}},
  \bibinfo{author}{\bibfnamefont{J.}~\bibnamefont{Peinke}}, \bibnamefont{and}
  \bibinfo{author}{\bibfnamefont{C.}~\bibnamefont{Renner}},
  \bibinfo{journal}{Phys. \ Rev. \ Lett.} \textbf{\bibinfo{volume}{84}},
  \bibinfo{pages}{5224} (\bibinfo{year}{2000}).

\bibitem[{\citenamefont{Ausloos and Ivanova}(2003)}]{ausloos_PRE}
\bibinfo{author}{\bibfnamefont{M.}~\bibnamefont{Ausloos}} \bibnamefont{and}
  \bibinfo{author}{\bibfnamefont{K.}~\bibnamefont{Ivanova}},
  \bibinfo{journal}{Phys. \ Rev. E} \textbf{\bibinfo{volume}{68}},
  \bibinfo{pages}{046122} (\bibinfo{year}{2003}).

\bibitem[{\citenamefont{Mizuno et~al.}(2007)\citenamefont{Mizuno, Takayasu, and
  Takayasu}}]{mizuno}
\bibinfo{author}{\bibfnamefont{T.}~\bibnamefont{Mizuno}},
  \bibinfo{author}{\bibfnamefont{M.}~\bibnamefont{Takayasu}}, \bibnamefont{and}
  \bibinfo{author}{\bibfnamefont{H.}~\bibnamefont{Takayasu}},
  \bibinfo{journal}{Physica A} \textbf{\bibinfo{volume}{382}},
  \bibinfo{pages}{187} (\bibinfo{year}{2007}).

\bibitem[{\citenamefont{Mandelbrot}(1963)}]{mandelbrot}
\bibinfo{author}{\bibfnamefont{B.~B.} \bibnamefont{Mandelbrot}},
  \emph{\bibinfo{title}{Journal of Business}}, vol.~\bibinfo{volume}{36}
  (\bibinfo{publisher}{University of Chicago Press}, \bibinfo{year}{1963}).

\bibitem[{\citenamefont{Gabaix et~al.}(2003)\citenamefont{Gabaix, Gopikrishnan,
  Plerou, and Stanley}}]{gabaix}
\bibinfo{author}{\bibfnamefont{X.}~\bibnamefont{Gabaix}},
  \bibinfo{author}{\bibfnamefont{P.}~\bibnamefont{Gopikrishnan}},
  \bibinfo{author}{\bibfnamefont{V.}~\bibnamefont{Plerou}}, \bibnamefont{and}
  \bibinfo{author}{\bibfnamefont{H.~E.} \bibnamefont{Stanley}},
  \bibinfo{journal}{Nature} \textbf{\bibinfo{volume}{423}},
  \bibinfo{pages}{267} (\bibinfo{year}{2003}).

\bibitem[{\citenamefont{Gopikrishnan et~al.}(1998)\citenamefont{Gopikrishnan,
  Meyer, Amaral, and Stanley}}]{gopi}
\bibinfo{author}{\bibfnamefont{P.}~\bibnamefont{Gopikrishnan}},
  \bibinfo{author}{\bibfnamefont{M.}~\bibnamefont{Meyer}},
  \bibinfo{author}{\bibfnamefont{L.~A.~N.} \bibnamefont{Amaral}},
  \bibnamefont{and} \bibinfo{author}{\bibfnamefont{H.~E.}
  \bibnamefont{Stanley}}, \bibinfo{journal}{European Physical Journal B: Rapid
  Communications} \textbf{\bibinfo{volume}{3}}, \bibinfo{pages}{139}
  (\bibinfo{year}{1998}).

\bibitem[{\citenamefont{Farmer and Lillo}(2004)}]{farmer}
\bibinfo{author}{\bibfnamefont{J.~D.} \bibnamefont{Farmer}} \bibnamefont{and}
  \bibinfo{author}{\bibfnamefont{F.}~\bibnamefont{Lillo}},
  \emph{\bibinfo{title}{Quantitative Finance}}, vol.~\bibinfo{volume}{4}
  (\bibinfo{publisher}{Routledge}, \bibinfo{year}{2004}).

\bibitem[{\citenamefont{Plerou et~al.}(2000)\citenamefont{Plerou, Gopikrishnan,
  Amaral, Gabaix, and Stanley}}]{plerou}
\bibinfo{author}{\bibfnamefont{V.}~\bibnamefont{Plerou}},
  \bibinfo{author}{\bibfnamefont{P.}~\bibnamefont{Gopikrishnan}},
  \bibinfo{author}{\bibfnamefont{L.~A.~N.} \bibnamefont{Amaral}},
  \bibinfo{author}{\bibfnamefont{X.}~\bibnamefont{Gabaix}}, \bibnamefont{and}
  \bibinfo{author}{\bibfnamefont{H.~E.} \bibnamefont{Stanley}},
  \bibinfo{journal}{Phys. \ Rev. E} \textbf{\bibinfo{volume}{62}},
  \bibinfo{pages}{R3023} (\bibinfo{year}{2000}).

\bibitem[{\citenamefont{Takayasu et~al.}({\natexlab{b}})\citenamefont{Takayasu,
  Watanabe, Mizuno, Ito, and Takayasu}}]{takayasu_bmodel}
\bibinfo{author}{\bibfnamefont{M.}~\bibnamefont{Takayasu}},
  \bibinfo{author}{\bibfnamefont{K.}~\bibnamefont{Watanabe}},
  \bibinfo{author}{\bibfnamefont{T.}~\bibnamefont{Mizuno}},
  \bibinfo{author}{\bibfnamefont{T.}~\bibnamefont{Ito}}, \bibnamefont{and}
  \bibinfo{author}{\bibfnamefont{H.}~\bibnamefont{Takayasu}},
  \bibinfo{note}{(in preparation)}.

\bibitem[{\citenamefont{Yamada et~al.}(2008)\citenamefont{Yamada, Takayasu, and
  Takayasu}}]{yamada_EPJB}
\bibinfo{author}{\bibfnamefont{K.}~\bibnamefont{Yamada}},
  \bibinfo{author}{\bibfnamefont{H.}~\bibnamefont{Takayasu}}, \bibnamefont{and}
  \bibinfo{author}{\bibfnamefont{M.}~\bibnamefont{Takayasu}},
  \bibinfo{journal}{The European Physical Journal B}
  \textbf{\bibinfo{volume}{63}}, \bibinfo{pages}{529} (\bibinfo{year}{2008}).

\bibitem[{\citenamefont{Takayasu et~al.}(2007)\citenamefont{Takayasu, Mizuno,
  and Takayasu}}]{takayasu_PhysicaA383}
\bibinfo{author}{\bibfnamefont{M.}~\bibnamefont{Takayasu}},
  \bibinfo{author}{\bibfnamefont{T.}~\bibnamefont{Mizuno}}, \bibnamefont{and}
  \bibinfo{author}{\bibfnamefont{H.}~\bibnamefont{Takayasu}},
  \bibinfo{journal}{Physica A} \textbf{\bibinfo{volume}{344}},
  \bibinfo{pages}{207} (\bibinfo{year}{2007}).

\bibitem[{\citenamefont{Ohnishi et~al.}(2004)\citenamefont{Ohnishi, Mizuno,
  Aihara, Takayasu, and Takayasu}}]{ohnishi_PhysicaA344}
\bibinfo{author}{\bibfnamefont{T.}~\bibnamefont{Ohnishi}},
  \bibinfo{author}{\bibfnamefont{T.}~\bibnamefont{Mizuno}},
  \bibinfo{author}{\bibfnamefont{K.}~\bibnamefont{Aihara}},
  \bibinfo{author}{\bibfnamefont{M.}~\bibnamefont{Takayasu}}, \bibnamefont{and}
  \bibinfo{author}{\bibfnamefont{H.}~\bibnamefont{Takayasu}},
  \bibinfo{journal}{Physica A} \textbf{\bibinfo{volume}{344}},
  \bibinfo{pages}{207} (\bibinfo{year}{2004}).

\bibitem[{\citenamefont{Takayasu et~al.}({\natexlab{c}})\citenamefont{Takayasu,
  Mizuno, and Takayasu}}]{takayasu_others}
\bibinfo{author}{\bibfnamefont{M.}~\bibnamefont{Takayasu}},
  \bibinfo{author}{\bibfnamefont{T.}~\bibnamefont{Mizuno}}, \bibnamefont{and}
  \bibinfo{author}{\bibfnamefont{H.}~\bibnamefont{Takayasu}},
  \bibinfo{note}{(in preparation)}.

\bibitem[{\citenamefont{Akaike}(1973)}]{akaike_1973}
\bibinfo{author}{\bibfnamefont{H.}~\bibnamefont{Akaike}}, in
  \emph{\bibinfo{booktitle}{Proc. 2nd International Symposium on Information
  Theory}} (\bibinfo{year}{1973}), p. \bibinfo{pages}{267}.

\bibitem[{\citenamefont{Schwarz}(1978)}]{schwarz_BIC}
\bibinfo{author}{\bibfnamefont{G.}~\bibnamefont{Schwarz}}, in
  \emph{\bibinfo{booktitle}{Annals of Statistics}}
  (\bibinfo{publisher}{Institute of Mathematical Statistics},
  \bibinfo{year}{1978}), vol.~\bibinfo{volume}{6}, p. \bibinfo{pages}{461}.

\end{thebibliography}
\end{document}